# IRANIAN MODAL MUSIC (DASTGAH) DETECTION USING DEEP NEURAL NETWORKS

**Danial Ebrat**
`Danial.et@aut.ac.ir`

**Farzad Didehvar**
`Didehvar@aut.ac.ir`

**Milad Dadgar**
`Milad.dadgar@aut.ac.ir`

Department of Mathematics and Computer science
Amirkabir University of Technology

## ABSTRACT

Music classification and genre detection are topics in music information retrieval (MIR) that many articles have been published regarding their utilities in the modern world. However, this contribution is insufficient in non-western music, such as Iranian modal music. In this work, we have implemented several deep neural networks to recognize Iranian modal music in seven highly correlated categories. The best model, BiLGNet, which achieved 92 percent overall accuracy, uses an architecture inspired by autoencoders, including bidirectional LSTM and GRU layers. We trained the models using the Nava dataset, which includes 1786 records and up to 55 hours of music played solo by *Kamanche*, *Tar*, *Setar*, *Reed*, and *Santoor* (Dulcimer). We considered Multiple features such as MFCC, Chroma CENS, and Mel spectrogram as input. The results indicate that MFCC carries more valuable information for detecting Iranian modal music (*Dastgah*) than other sound representations. Moreover, the architecture inspired by autoencoders is robust in distinguishing highly correlated data like *Dastgah*s. It also shows that because of the precise order in Iranian *Dastgah* Music, Bidirectional Recurrent networks are more efficient than any other networks that have been implemented in this study.

## 1. INTRODUCTION

Music classification and genre detection are among the most popular and essential topics that have attracted scientists for years as they can be used for various tasks such as recommender systems or music tagging. In western music, which can be considered tonal music, comprehensive datasets are introduced for scientific studies that have led to remarkable results. Consequently, they are now employed in plenty of music platforms to offer a better music experience for a user. Unfortunately, such satisfactory datasets are rare in a vast amount of non-western music. Some non-western traditional music, such as Iranian, Turkish, and Arabic, are among those music cultures that are not well-known and have a significant gap to fill by MIR researchers.

In Iranian music, the number of research on MIR topics has been less than 20 in the last 20 years. Additionally, many of these works are written in Persian and have never been published in an international format. Furthermore, we can observe that each has been performed using an exclusive dataset, and there are no two studies available that have been performed on the same dataset. Consequently, they are incomparable, and nobody could say which method is better or achieves higher accuracy because of the lack of a suitable benchmark. Also, due to the complexity of Iranian traditional music, most of those datasets are not reliable in terms of richness and variousness. Despite western music, Iranian music is modal music consisting of various highly correlated modes or classes known as *Dastgah*. This correlation is too high to be recognized by most people, and an experienced vocalist or musician can perform all of them pleasingly within a one-minute music piece.

In previous attempts, various methods have been utilized to propose a method for Iranian *Dastgah* detection. Nevertheless, only a few used deep learning methods to categorize highly correlated classes. Furthermore, considering the fundamental differences between Iranian modal music and western genre classification, proven methods and architectures might not be entirely appropriate to detect a *Dastgah* accurately. Therefore, in this research, we sought the appropriate architecture for a deep learning model to detect Iranian *Dastgah* music considering Iranian music structure. We proposed BiLGNet, a deep neural network inspired by autoencoder architecture to distinguish highly correlated modes. We trained the model on the Nava dataset, a comprehensive Iranian music dataset in terms of completeness and variety. Moreover, considering the importance of sequential patterns and discipline in Iranian music, we utilized bidirectional LSTM and GRU networks to capture sequential patterns. Furthermore, finding the best representation of a sound that contains the most valuable information about Iranian modal music structure was another objective of this research. Thus, we used several music representations as inputs to the models, where the best results were achieved when we used MFCC as the input.

## 2. IRANIAN MODAL MUSIC (DASTGAH)

In the formation of Iranian modal music, small parts join together to create a specific mode (*Dastgah*). In other words, the formation of Iranian traditional music has a bottom-up structure, including several steps. Each mode contains multiple correlated parts known as *Gushe*, and each *Gushe* is created from several segments that can be used to identify a specific one.

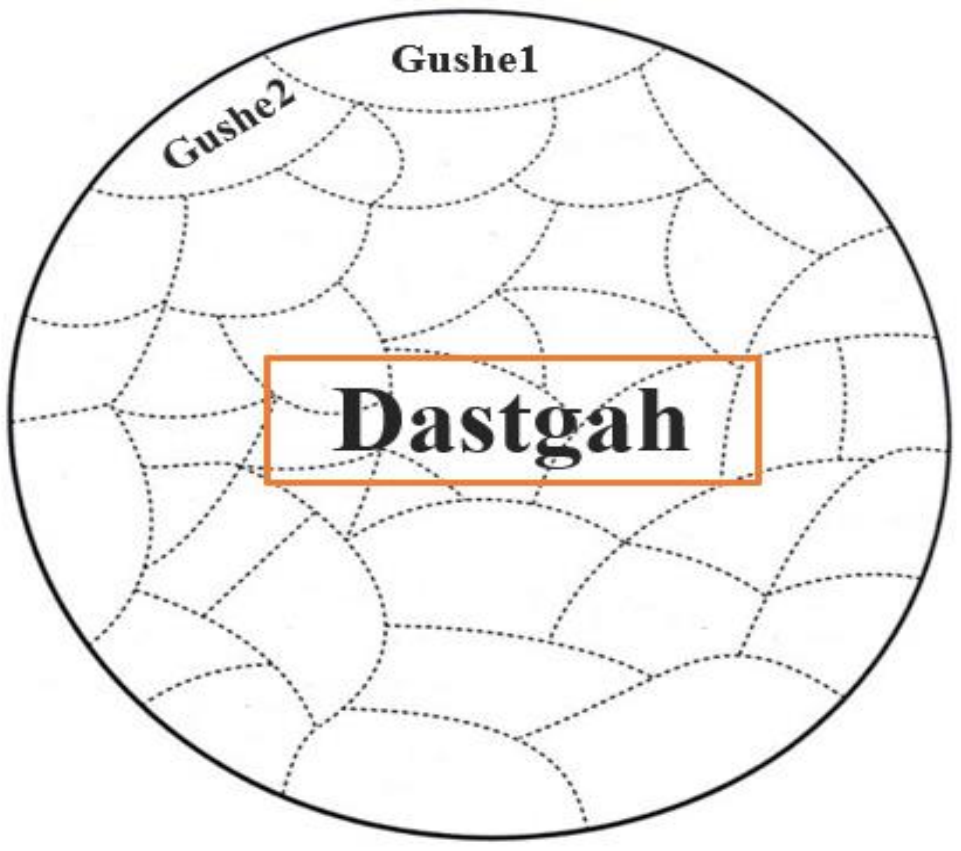

**Figure 1**. Formation of *Dastgah*

The formation of *Gushe* has several steps. In the first stage, a series of notes create a melody, and those melodies create a sentence. These sentences are a unique component of *Gushe*s, where they can serve as an identifier because of their characteristics. If a *Gushe* does not have an identification sentence, it can be recognized by its other features, such as rhythm, poem structure, or melody lines. In addition, each *Gushe* can appear in multiple *Dastgah*s to give the musician the freedom to use them as a bridge in order to perform several modes in one piece of music. This attribute of the *Gushe*s is the main reason for the high correlation among all modes that that a minimum of 43% and a maximum of 73% similarity between two arbitrary modes can be observed [1]. The number of well-known Gushes is approximately 250 in all *Dastgah*s. However, the number of all *Gushe*s is much more than the proposed number.

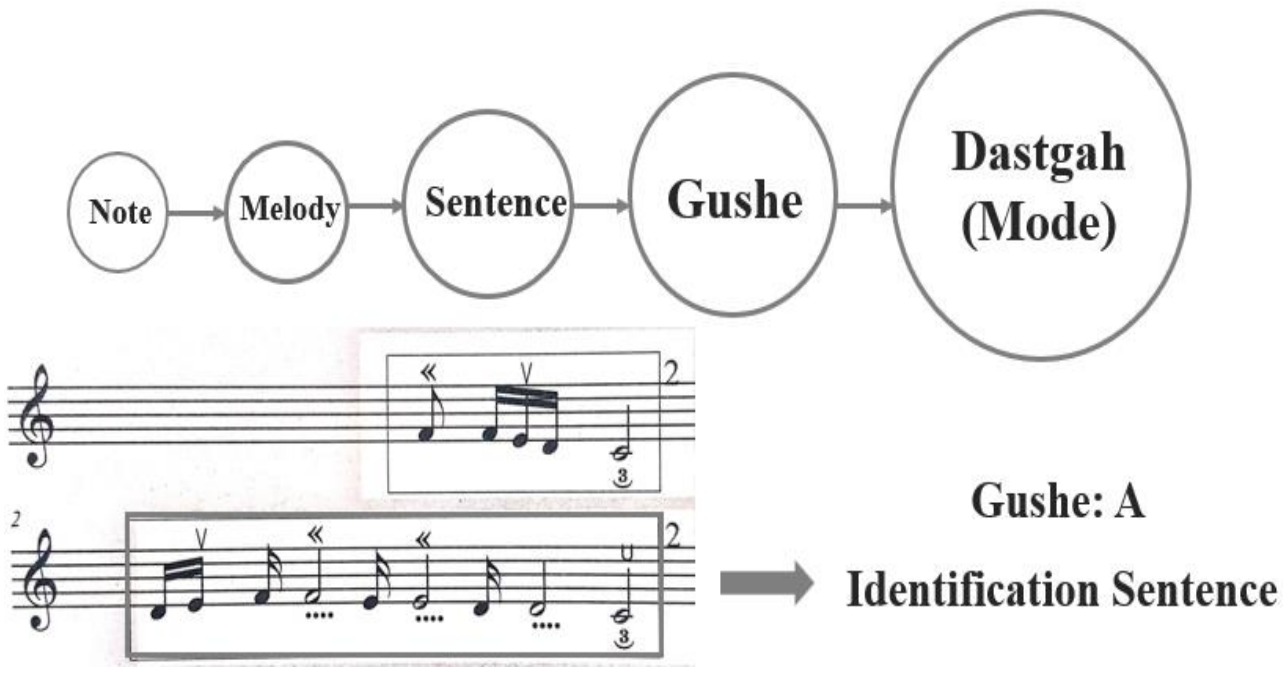

**Figure 2**. Formation of the Iranian *Dastgah* music

In theory, each octave in the Iranian traditional music scale consists of 24 equal notes considering quartertones. Quartertones are not exactly half of the semitones and can reside anywhere between two semitones. However, in practice, based on Safi al-Din al-Urmawi's theory, each octave consists of 17 performable notes with unequal intervals.

*Dastgah* is a set of multi-modes, and each *Dastgah* includes the entire range of audio suitable for the human voice that can be started at any tonic note. The range of singing for humans is considered to be two and a half octaves in Western music, and this distance is equal to six *Dong*s in Iranian music, where each octave is divided into two *Dong*s. Considering an identical starting note for each *Dastgah*, we can see that every two arbitrary *Dastgah*s are similar in pitch classes, at least in one of the *Dong*s (approximately half of the octave). This similarity is so high that shifting a note by a quartertone would entirely change the mode (*Dastgah*). As an example, by increasing the third note (the third distance from the tonic note) of one mode (*Shur*) by a semitone, without any other change in the intervals of the pitch class, another mode (*Homayoun*) will have appeared. Also, another mode (*Segah*) would appear by reducing the third note by a quartertone. Therefore, multiple theories have been formed that propose a diverse number of classes for Iranian *Dastgah* music. Due to the high similarity between modes, some theories compartmentalize *Dastgah* into five classes, though in most theories, *Dastgah* has been categorized into seven classes (*Shur*, *Mahur*, *Homayoun*, *Chahargah*, *Rastpanjgah*, *Nava*, *Segah*), which is the base assumption of this paper. Moreover, because of the completeness of Iranian vocal music, in some theories, vocals have their additional categorization, which is five classes. Thus, in some theories, considering vocals and instruments, 12 classes have been proposed for the formation of Iranian *Dastgah* music.

## 3. RELATED RESEARCH

Many studies have been done especially associated with music from the beginning of the connection between art and artificial intelligence. However, as mentioned before, contributions to Iranian music have not been considerable in the past years. This section attempts to review all the studies (in English and Persian) that have been done on Iranian modal music and highlight the most influential parts. [2] introduced the first attempt to detect *Dastgah*. He reviewed pitches and repetitive melodies at each scale and compared them to other scales under the supervision-based method. An average error rate of 28.2% is obtained on performing the method on one of the *Dastgah*s. [3] used pitch profile and CQT conversion to determine the similarity where the average accuracy of 71% was reported. [4] was the first research that used an artificial neural network, and its purpose was to identify only one *Dastgah* (*Mahur*) from five others (considering five classes for categorization). She proposed an RBF network trained on the top 20 peaks extracted using FFT. The model's accuracy in detecting one specific *Dastgah* (*Mahur*) is estimated to be 70%. [5] also used the RBF network and the SVM classifier trained on the pitch, MFCC, and Spectral Centroid. The accuracy is estimated between 65% and 95% for various *Dastgah*s. [1] utilized a type-2 fuzzy logic, and the overall accuracy has been reported to be 85%. One of the notable results of this research is the measurement of the similarity between *Dastgah*s. According to this study, at least 43% and at most 73% similarity is observable in every two arbitrary *Dastgah*s. [6] objective was to distinguish *Gushe*s of one specific *Dastgah* (*Shur*). He applied FFT to extract five repetitive melodies from each *Gushe* to train a neural network. The average error rate is estimated at 10.87%. In [7], despite previous studies, the frames' length was considered dynamically with respect to the beginning of the spikes, and music pieces were used without segmentizing. This research introduced a method that achieved 93% average accuracy in detecting *Dastgah*s. He reported that the model achieved 93% average accuracy in detecting *Dastgah*s.

However, the model was trained on a dataset that each *Dastgah* was started from a specific tonic note. Therefore, the result is not reliable by changing the starting note. In [8], 330 pieces have been studied, including 180 *Gushe*s, all played with the dulcimer and collected from a training album. This research was the first attempt at using the hidden Markov model considering Zero crossing rate, Spectral Rolloff, Spectral Flux, and MFCC. The average accuracy of the system for detecting seven *Dastgah*s is estimated to be 74.5%. [9] was the first research that considered ensemble performance. He used 143 music pieces performed solo by five instruments as well as ensemble performances. This research introduces a Gaussian Mixture Model to detect the similarities using spectrogram and chroma features where the accuracy rates of 90.11% and 80.2% were achieved, respectively, using Manhattan distance. [10] trained an MLP model using the top 20 peaks of spectrograms extracted by FFT. For this research, 348 music pieces were performed solo by reed, violin, and vocal, where the accuracies obtained were 65%, 72%, and 56%, respectively. In [11], the Nava dataset is introduced for *Dastgah* classification and instrument recognition tasks. In terms of the instrument, *Dastgah* and artist Nava have enough comprehensiveness and variety compared to previous studies. Thus, we employed the Nava dataset throughout this research. More detail about the dataset has been explained in the next section. Also, the best accuracy obtained on the Nava dataset for the *Dastgah* classification is 34%. In [12], AzarNet has been proposed, which is the first Deep learning attempt for *Dastgah* recognition. The output of STFT has been used as an input for the model that contains CNN and GRU networks. It is worth stating that their dataset contains 1137 music tracks played solo with two instruments (violin and reed). For an average of all seven classes, the overall f1 score of AzarNet on the test set was 86.21%. [13] has attempted to detect *Gushe* instead of *Dastgah*. For the simplicity of the goal, only one of the *Dastgah*s (*Shur*) has been studied, and the method can be extended to other *Dastgah*s. for this purpose, he segmentized music pieces regarding the start and ending point of melody lines and identification sentences rather than equal segments. In the proposed method, first, the notes have been extracted using a periodogram, and a database of the note-time matrix has been created for each identification sentence, which can be utilized to calculate the similarity of new segments. Unfortunately, no accuracy of the mentioned model has been announced. In [14], one of the impressive results is the construction of an extensive and comprehensive dataset that includes more than 2000 music pieces of solo and ensemble performances of various Iranian instruments. Unfortunately, this dataset is not publicly available yet. Moreover, he trained a CNN model using the raw signals, and the estimated average accuracy of 86% is reported for *Dastgah* detection. In [15], the most recent research on this domain, Iranian traditional vocal music pieces have been categorized into 12 modes (considering five additional classes for vocal). They extracted multiple audio features such as pitch frequency histogram and MFCC histogram to train an MLP model that achieved 74.2% average accuracy using ten times operating on 16 vocal samples of the test set. Finally, [16] is the only research on *Dastgah* detection subject that a non-Iranian researcher has done. This study concentrates on monophonic audio recordings using Markov Models, where the sequence of intervals is calculated from quantized pitch data (estimated from audio) to create a Markov model. Consequently, the classification is performed by finding the closest match between the Markov matrix of the file and the matrices computed for each *Dastgah*. A leave-one-out evaluation strategy was implemented for the evaluation procedure on a dataset comprised of 73 files, and an accuracy of 0.986 was observed using the Bhattacharyya likelihood distance. Although the method is powerful, it is needed to be performed on a much bigger dataset.

## 4. METHODOLOGY

Previous methods and accuracy achieved in detecting Iranian modal music show that although machine learning techniques present reasonable and reliable approaches, they might not be strong enough to reach the perfect exactness due to the complexity of the problem. In theory, Iranian modal music proposes a definable structure for its nature; However, practically, this structure is entirely adjustable, and musicians can perform this music regarding their distortions and ingenuity to create new music. As a result, improvising is an inseparable part of Iranian *Dastgah* music. Therefore, the complexity of *Dastgahs* is too high that only well-experienced musicians can identify them from audio accurately, and this accuracy derives from their ear's adaptation to different modes, not from theoretical reasons. Therefore, we trained several deep neural networks to detect Iranian *Dastgah* from audio music in a way that musicians perceive. In order to feed the data to a deep neural network, we have performed several preprocessing stages, which will be discussed in what follows. Furthermore, this section presents the models that we implemented in this paper, where the best one, BiLGNet, has been described in more detail.

### 4.1 Preprocessing

We implemented several steps to create suitable input for deep learning models. These steps are explained briefly in the following paragraphs. Also, Figure 3 illustrates the schematic diagram of the stages.

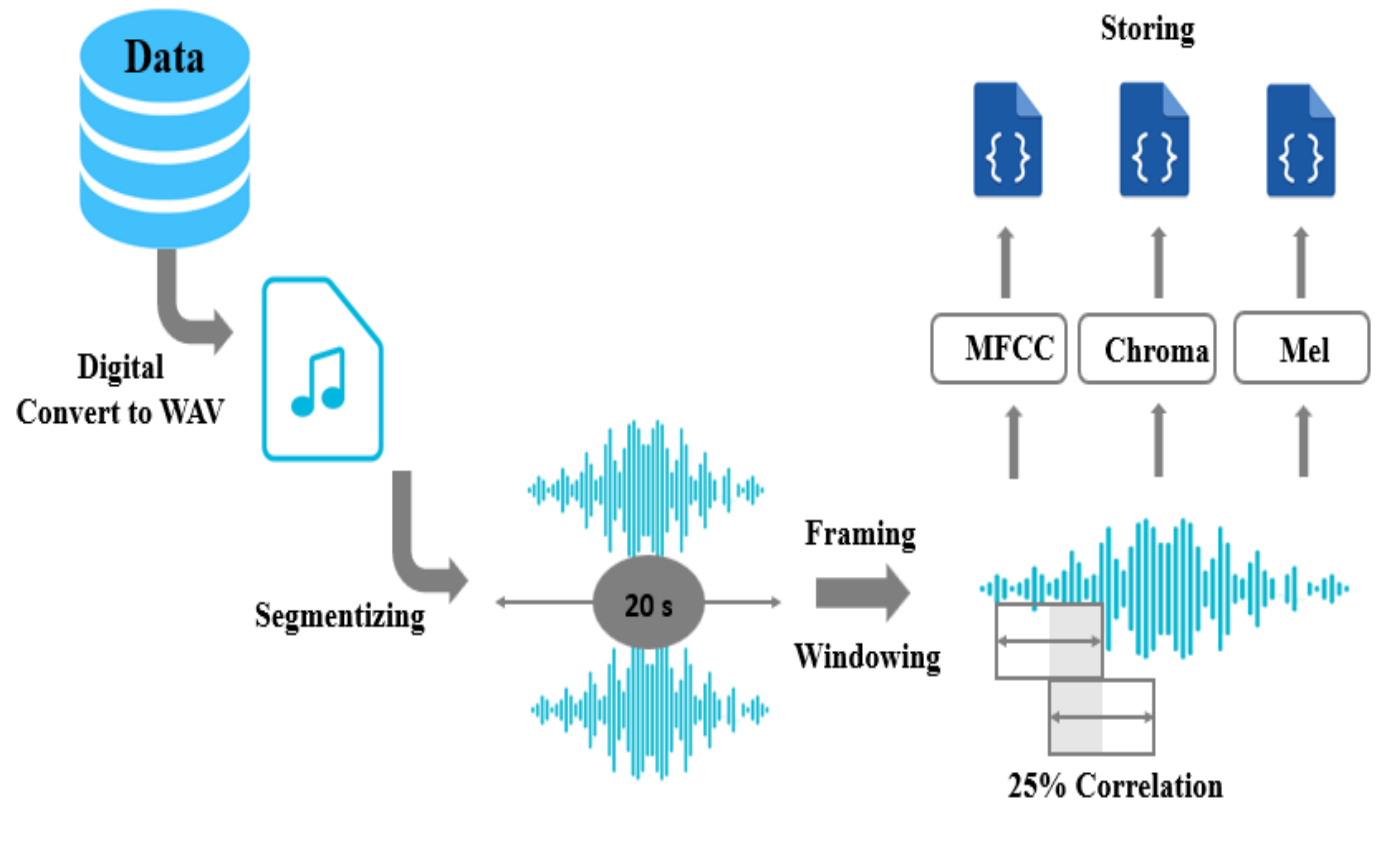

**Figure 3.** Preprocessing procedure

#### 4.1.1 Stage 1: Initial Conversion

In the first stage, in order to process music, the analog signal has to be converted into a digital signal using sampling (22050 Hz) and Quantization (16 bit). Furthermore, we converted the music files into the WAV format since we used the Librosa library [17] through the preprocessing stages.

#### 4.1.2 Stage 2: Segmentizing

In this stage, we divided every music piece into distinguished segments with equal lengths without considering any correlation. The average time to detect a music mode by a professional Iranian musician is more than 10 seconds approximately since the correlation between modes is too high to be recognized accurately in the short term. Thus, we considered 20 seconds long for the length of segments to disregard unwanted correlations and increase the accuracy. Also, we should note that models trained on 10 and 15 seconds segments had a minor accuracy decrease.

#### 4.1.3 Stage 3: Applying framing and windowing

The most common procedure for extracting information from a given signal is applying the Fourier transform and its variations. This process will convert the signal into its constituent components. The signal has to be broken into small frames in order to detect the time and the sequential order of constituent components, which is called framing. The duration of the frames should be greater than the detectable tone for humans, which is approximately ten milliseconds. For this goal, we set the framing length to 2048, which means 2048 sample points in each frame (about 90 milliseconds). In order to avoid spectral leakage, we applied the Hann windowing function to each frame to smooth the unwanted spikes at the beginning and end of the frames. At last, we considered 25% correlation in the framing procedure to avoid data loss after the windowing function.

#### 4.1.4 Stage 4: Applying Fourier transform

By applying STFT on each frame, we extracted various features and saved them individually in order to use them as the input for our models. Mel spectrogram, Chroma CENS, and MFCC are the features we extracted through this stage. Regarding the existence of quartertones in Iranian music arrangements, we considered 24 coefficients and bins for MFCC and Chroma CENS. Moreover, 128 Mel bands were used for the Mel spectrogram to capture the subtleties of Iranian music. We assessed these numbers through various implementations, and we used different numbers for mentioned music representations to find the most appropriate ones.

### 4.2 Models

To conquer the high similarity problem between various modes, the architecture used in all models is inspired by the autoencoder architecture. In some cases, we combined them with residual architecture. We implemented numerous deep learning models throughout this research, and the most compelling ones are mentioned in Table 4. In different parts of the models, we utilized various networks such as CNN, LSTM, GRU, Bidirectional LSTM, Bidirectional GRU, and hybrid networks in order to reach the most suitable model possible. In addition, the implementation of the models is accessible through the research repository GitHub[1]. The following section will discuss comprehensively the best model which attained the highest total accuracy.

### 4.3 BiLGNet

The best model in terms of accuracy is a deep neural network composed of Bidirectional LSTM, Bidirectional GRU, and fully connected layers. The network starts with three Bidirectional LSTM layers, where each one is followed by batch normalization, as shown in figure 4. The number of neurons in the first layer equals 128, and for the following layers, each holds half of the neurons in the previous layer. After mapping the output of Bidirectional LSTM to a latent space consisting of a Bidirectional GRU with 16 neurons, the results are fed to three Bidirectional GRU layers, where each one is followed by a batch normalization. Moreover, the number of neurons in each layer is double the number of neurons in the previous one. Because of the necessity of time series patterns and temporal features in Iranian music and the precise order of the elements in both time and counter time order, Bidirectional RNN-based networks outperform one-way networks and CNNs. In the next step, the output of the last Bidirectional GRU is sent to a fully connected layer which plays the role of a bottleneck with 16 neurons and a dropout layer to avoid overfitting. The bottleneck has shown promising results in recognizing acoustic data [18] compared to models that we implemented without a bottleneck. Finally, the number of neurons included in the output layer, which is another fully connected layer, is equal to seven, where each one represents a *Dastgah* in Iranian music. A brief diagram of BiLGNet, including layers and the number of neurons in each one, is shown in figure 4.

### 4.4 Network Parameters

In the training process, we applied multiple activation functions such as Tanh and Sigmoid for the Bidirectional Recurrent networks (default activation functions in Keras), ReLu in the bottleneck, and the Softmax for the output layer since each sample belongs to only one *Dastgah*. Furthermore, The Adam optimizer has been utilized with the default learning rate of 0.001. We implemented the learning rate dynamically, so it would be decreased by 0.7 if the accuracy enhancement for evaluation data cannot reach the considered value for every seven epochs. Moreover, we employed the sparse categorical cross-entropy as the loss function to adjust the network variables. Finally, we set the batch size to 32 for the learning process.

[1] https://github.com/danialebrat/Dastgagh_classification

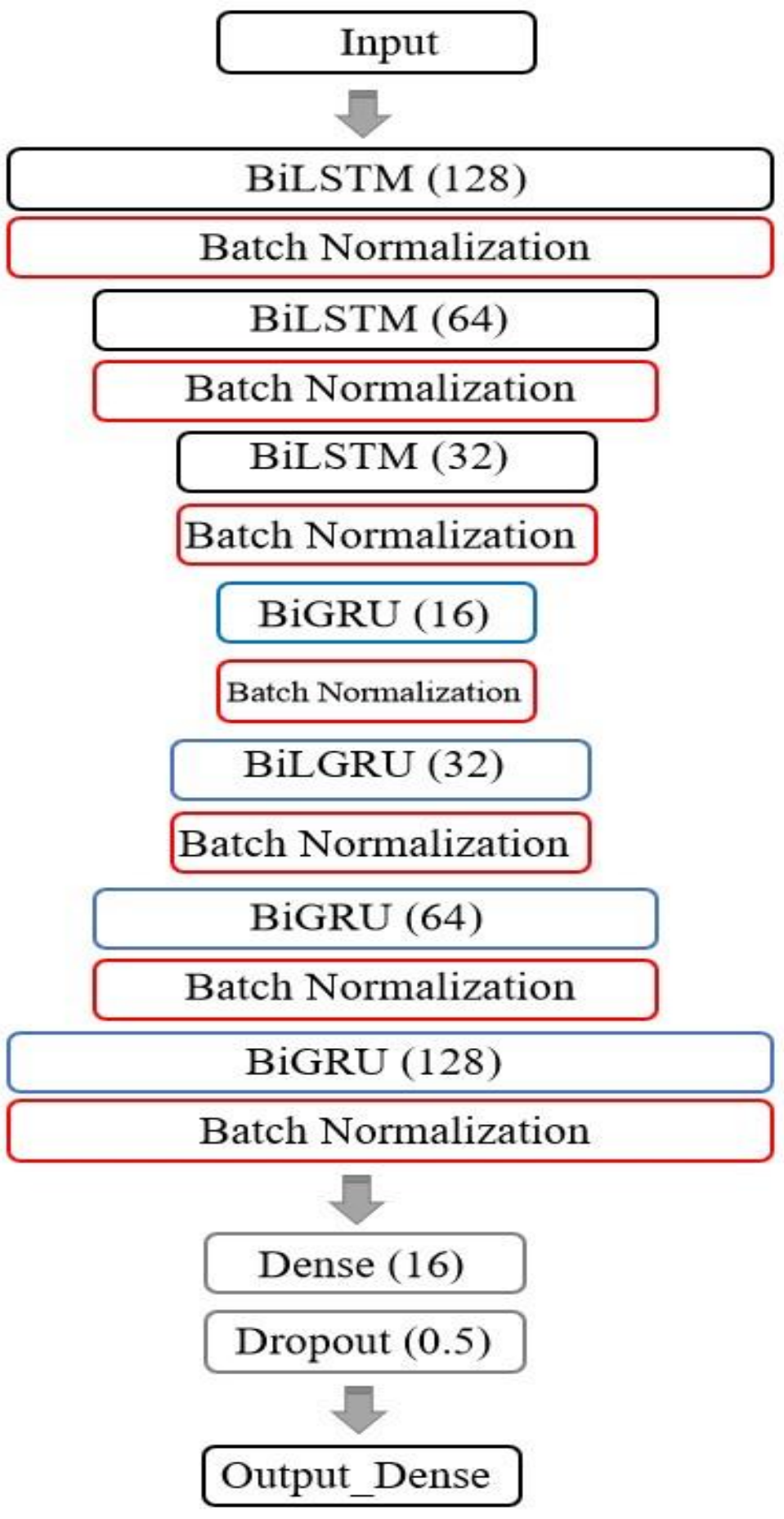

**Figure 4.** BiLGNet Model

## 5. DATASET

In this research, we used the Nava dataset [11] which includes 1786 records and up to 55 hours of music played solo by 40 various artists with five known Iranian instruments (*Kamanche*, *Tar*, *Setar*, *Reed*, and *Dulcimer*). The following tables show the distribution of instruments and music records in each *Dastgah*. Moreover, each music piece belongs only to one of the *Dastgah*s. Regarding the completeness of this dataset, the Nava dataset can be considered a starting point for creating a benchmark in Iranian *Dastgah* detection.

| | **Dulcimer** | **Reed** | **Setar** | **Tar** | **Kamancheh** |
|---|---|---|---|---|---|
| **Shur** | 88 | 47 | 53 | 57 | 34 |
| **Mahur** | 50 | 48 | 82 | 42 | 50 |
| **Chahargah** | 71 | 57 | 67 | 42 | 34 |
| **Homayoun** | 65 | 69 | 41 | 50 | 28 |
| **Segah** | 73 | 50 | 39 | 67 | 23 |
| **Nava** | 53 | 57 | 40 | 51 | 41 |
| **Rastpanjgah** | 38 | 66 | 42 | 39 | 32 |

**Table 1**. Number of music pieces performed by each instrument in *Dastgahs*

| **instrument** | Dulcimer | Reed | Setar | Tar | Kamancheh |
|---|---|---|---|---|---|
| **Duration (hour)** | 17.1 | 8.48 | 12.01 | 9.23 | 8.16 |

**Table 2**. Duration of each instrument

| **Dastgah** | **Duration (hour)** | **Number of records** |
|---|---|---|
| Shur | 8.33 | 279 |
| Mahur | 8.42 | 272 |
| Chahargah | 9.22 | 271 |
| Homayoun | 8.12 | 253 |
| Segah | 8.09 | 252 |
| Nava | 6.4 | 242 |
| Rastpanjgah | 6.39 | 217 |

**Table 3**. Duration of each *Dastgah* and number of records

## 6. RESULT AND DISCUSSION

After the preprocessing, we obtained more than 9,000 samples which we considered 90% of them as training data, 5% (451 samples) for the testing procedure, and 5% for evaluating the models. Table 4 shows the total accuracy achieved by each model, including the input format of the model. As we can see, BiLGNet has gained the best result. The prominent point is the better performance of MFCC compared to other representation formats, which shows that MFCC contains more trainable information about the sequential patterns hidden in Iranian music. Furthermore, it also shows that MFCC can carry data other than just timbre information, as shown in other studies [19]. The results demonstrate that in an identical configuration, all the models achieved higher accuracy (approximately between 3% to 7%) if they received MFCC as the input.

| **Model** | **Feature** | **Total Accuracy (%)** |
|---|---|---|
| Residual Conv2D | MFCC | 68 |
| Residual Conv1D | MFCC | 79 |
| Conv1D + BiGRU | MFCC | 88 |
| Residual BiLGNet | MFCC | 88 |
| BiLGNet | Chroma CENS | 85 |
| BiLGNet | MFCC | 92 |

**Table 4.** Results

Chroma CENS was the initial guess that could be the most appropriate representation to train various models considering solo performances in music pieces. Nonetheless, although the results of training models on Chroma CENS are noticeable compared to previous studies, MFCC performs better in all cases.

Moreover, as mentioned before, we considered different numbers of coefficients and bins for the representation methods. The results show a noticeable accuracy increase with the enhancement of Mel bands while extracting Mel spectrograms. In our opinion, considering quartertones that can reside anywhere between two semitones, a low number of Mel bands cannot capture all the information. We realized this in the initial executions that we performed on a small proportion of the dataset we built for time-saving purposes. Therefore, considering initial executions, Mel bands should be more than 130. Nevertheless, MFCC and Chroma CENS's superiority was also evident in the early executions performed on a small proportion of the dataset. On top of that, the volume of the Mel spectrogram preprocessing file was too heavy to be used for learning procedures considering available facilities.

### 6.1 Evaluation of the best model

We have evaluated the model by confusion matrix and classification report. Figure 5 illustrates the confusion matrix, which shows the classification of the test dataset for each mode. As we can see, the model has succeeded in detecting all 83 samples of *Chahargah*'s mode precisely. Furthermore, the classification report, including the F1 score for each mode, is shown in Table 5.

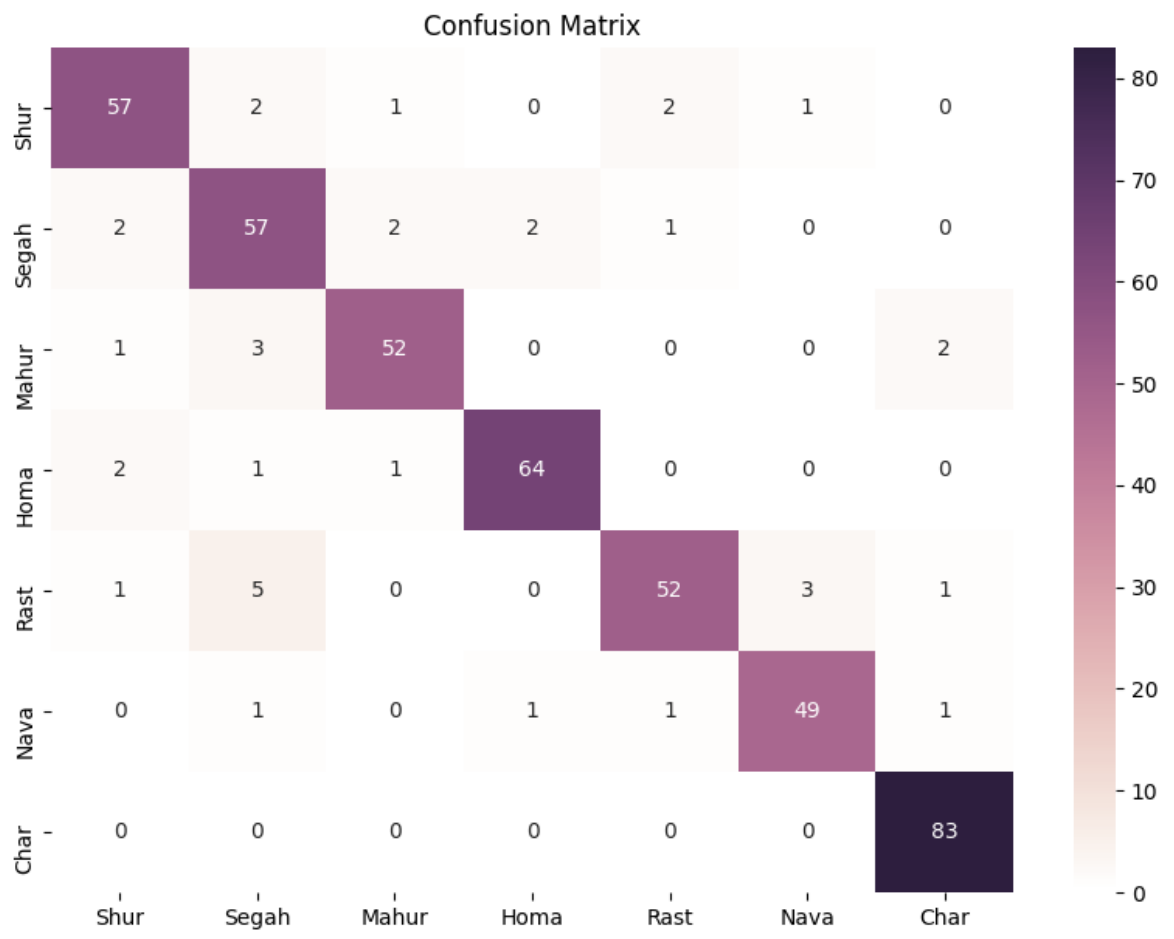

**Figure 5.** Confusion matrix

| Mode | Precision | Recall | F1-Score | Support |
|---|---|---|---|---|
| Shur | 0.90 | 0.90 | 0.90 | 63 |
| Segah | 0.89 | 0.83 | 0.86 | 69 |
| Mahur | 0.90 | 0.93 | 0.91 | 56 |
| Homayoun | 0.94 | 0.96 | 0.95 | 67 |
| Rastpanjgah | 0.84 | 0.93 | 0.88 | 56 |
| Nava | 0.92 | 0.92 | 0.92 | 53 |
| Chahargah | 1.00 | 0.95 | 0.98 | 87 |
| Total Accuracy | 0.92 | | | 451 |
| Average Macro | 0.91 | 0.92 | 0.91 | 451 |
| Average weighted | 0.92 | 0.92 | 0.92 | 451 |

**Table 5.** Classification report

## 7. CONCLUSION

This research attempted to categorize Iranian music into seven highly correlated modes or *Dastgah*. We implemented several deep learning models where BiLGNet achieves a total accuracy of 92% that outperforms previous deep learning-based approaches. The results indicate that the MFCC carries more valuable information than other features studied through this research. Each mode's advent (*Dastgah*) has a precise sequential pattern dependent on music intervals and the frequency ratio between notes rather than the tonal information. Furthermore, each part of a mode, *Gushe*, can appear in multiple modes. For recognizing the current state of a particular one, the previous *Gushe* or the one that arises after has to be identified. Consequently, Bidirectional RNN-based networks illustrate a better performance due to their power to capture time-series patterns than other models using CNNs or hybrid models. In addition, we show that the models inspired by the autoencoder architecture acquire the highest accuracy comparing the same model without mapping the data in a small latent space.

## 8. FUTURE WORKS

Unfortunately, the lack of a proper dataset for Iranian music has been a great challenge for years. We can see that all the previous attempts have been implemented on a customized exclusive dataset. Therefore, there are no two studies on a common dataset to observe a standard comparing procedure. Besides, only a few of those datasets are reliable in terms of richness. The variety of artists and instruments in the Nava dataset is an appropriate starting point for future works that could create a benchmark. Furthermore, in our opinion, the cooperation of online music platforms, artists, and scientists in this field can be valuable in creating a more suitable dataset, including ensemble performances, in which the samples can be labeled by *Gushe* instead of *Dastgah*. Building such a dataset will modify the research methods associated with Iranian music, and the mode can be identified in a bottom-up structure in the exact way that appears in Iranian music.

Moreover, the differences between *Dastgah*s are recognizable when they start at the same tonic note. As mentioned before, tonal information is not promising in detecting a *Dastgah*; However, mapping all the present notes of each music piece in a specific octave scale could expose the dissimilarities in a more comprehensible way.

At last, regarding the similarity of the Iranian music to other music, such as Turkish, Arabic, and Azerbaijanian music, BiLGNet and the proposed method can also be applied to other similar music.